\documentclass{moriond}

\def\be{\begin{equation}}
\def\ee{\end{equation}}
\def\bea{\begin{eqnarray}}
\def\eea{\end{eqnarray}}

\newcommand{\Photo}{}

\begin{document}
\vspace*{4cm}
\title{Charm total cross sections with nonuniversal fragmentation treatment}

\author{ A. Geiser (speaker), Y. Yang }
\address{DESY Hamburg, Notkestrasse 85, D-22607 Hamburg, Germany}
\author{ S. Moch, O. Zenaiev}
\address{Hamburg University, Luruper Chaussee 149, D-22761 Hamburg, Germany}

\maketitle\abstracts{
Total charm-pair cross sections in $pp$ collisions are interesting because 
they can be calculated to NNLO in QCD without any reference to fragmentation
effects. On the other hand, the fiducial differential charm cross sections 
from which the total cross sections must be extrapolated are currently 
known to NLO+NLL at most (e.g. FONLL), and must be treated for known effects of 
nonuniversal charm fragmentation. A new procedure using the FONLL framework 
as input for an empirical parametrization of the data in both shape and 
normalization, with all its parameters actually fitted to data, is used to 
derive so-called data-driven FONLL (ddFONLL) 
parametrizations which can be used to extrapolate the differential 
cross sections to total cross sections with minimal bias. 
This includes an empirical 
treatment of all known non-universal charm fragmentation effects, 
in particular for the baryon-to-meson ratio as a function of 
transverse momentum.
The total charm-pair cross sections obtained in this way, which supersede 
all previous ones obtained using the assumption of charm-fragmentation 
universality, are consistent with NNLO predictions, and allow first studies 
of their sensitivity e.g. to the charm-quark mass and/or the NNLO gluon PDF at 
very low transverse momentum fractions $x$.}

\section{Introduction}

The theory of Quantum-Chromo-Dynamics (QCD) is a well-established part of the 
Standard Model of particle physics, which describes most of the processes
occurring in $pp$ collisions at LHC rather well. Predictions for charm 
production are particularly challenging since, due to the closeness of 
the charm quark mass, 
$m_c \sim 1.5$ GeV, to the nonperturbative QCD scale, $\Lambda_{QCD} \sim 0.3$
GeV, the convergence of the 
perturbative series is slow, resulting in large theoretical uncertainties. 
In addition, nonperturbative effects enter the hadronization of charm quarks 
into charm hadrons, the entities actually measurable in detectors.
Recently, it was established\,\cite{nonuni1,nonuni2,nonuni3,nonuni4} 
(Fig.~\ref{fig:ftilde}) 
that the corresponding charm fragmentation effects
are non-universal, i.e. can not simply be transferred from $e^+e^-$ and $ep$ 
measurements to $pp$ measurements and may be dependent on kinematics. 
Finding an appropriate treatment of this non-universality
is one of the central themes of this contribution.  

Measuring the total cross section for charm-quark-antiquark pair production,
$\sigma_{c\bar c}^{\mathrm{tot}}$, almost\,\footnote{Hadron states containing a 
$c\bar c$ pair or more than one $c$ quark contribute only about 1\% to the 
total charm cross section and are hence neglected here.} 
identical to the total summed and integrated cross section for 
the prompt production of charm hadrons $H_c$ containing a single charm quark $c$
(not $\bar c$), 
$\sigma_{H_c}^{\mathrm{all}}$, 
is of particular interest since the corresponding theoretical predictions are 
currently the only $pp$ charm cross sections calculable at 
next-to-next-to-leading order (NNLO) in QCD, and do furthermore not depend 
upon charm fragmentation. As we will demonstrate, they can thus be used 
to derive constraints on genuine QCD parameters such as the charm-quark 
mass $m_c$ and the proton partons density functions (PDFs), at NNLO. 
For such measurements, differential $H_c$ distributions measured in limited 
kinematic ranges (fiducial cross sections) and for a restricted set of $H_c$ 
final states need to be extrapolated to the total cross section, under certain
constraints concerning their shape and normalization in unmeasured regions.
Furthermore, the possibly phase space dependent relative contributions of 
different $H_c$ final states (nonuniversality) need to be accounted for.  

In this contribution we will describe how such shape, normalization
and nonuniversality constraints can be obtained in an entirely data driven 
way, based on a 
theory-inspired parametrization of the multi-differential cross sections,
and fully accounting for charm nonuniversality effects, without attempting 
to rely on any particular nonuniversal charm fragmentation model.
Since space is limited, we will extensively refer to a previous 
writeup\,\cite{EPS23} where applicable, and defer some other discussions to 
yet another writeup to appear soon\,\cite{DIS24}.      

\section{FONLL vs. ddFONLL calculations}

The highest order calculations available so far for differential 
charm-production cross sections are based on the massive next-to-leading 
order (NLO) plus massless next-to-leading-log (NLL) approach, also known as 
general mass variable flavour number scheme, one example of 
which, exploited here, is the FONLL\,\cite{FONLL} approach. 
By construction, the NLL part of this calculation becomes significant only 
at high values of charm transverse momentum ($p_{Tc} \gg m_c$), such that, 
since the total cross section is dominated by the low $p_T$ contribution,
NLO massive fixed order calculations without resummation could also be used 
instead for this purpose.     
    
\begin{figure}
\begin{minipage}{0.65\linewidth}
\centerline{\includegraphics[width=0.95\linewidth]{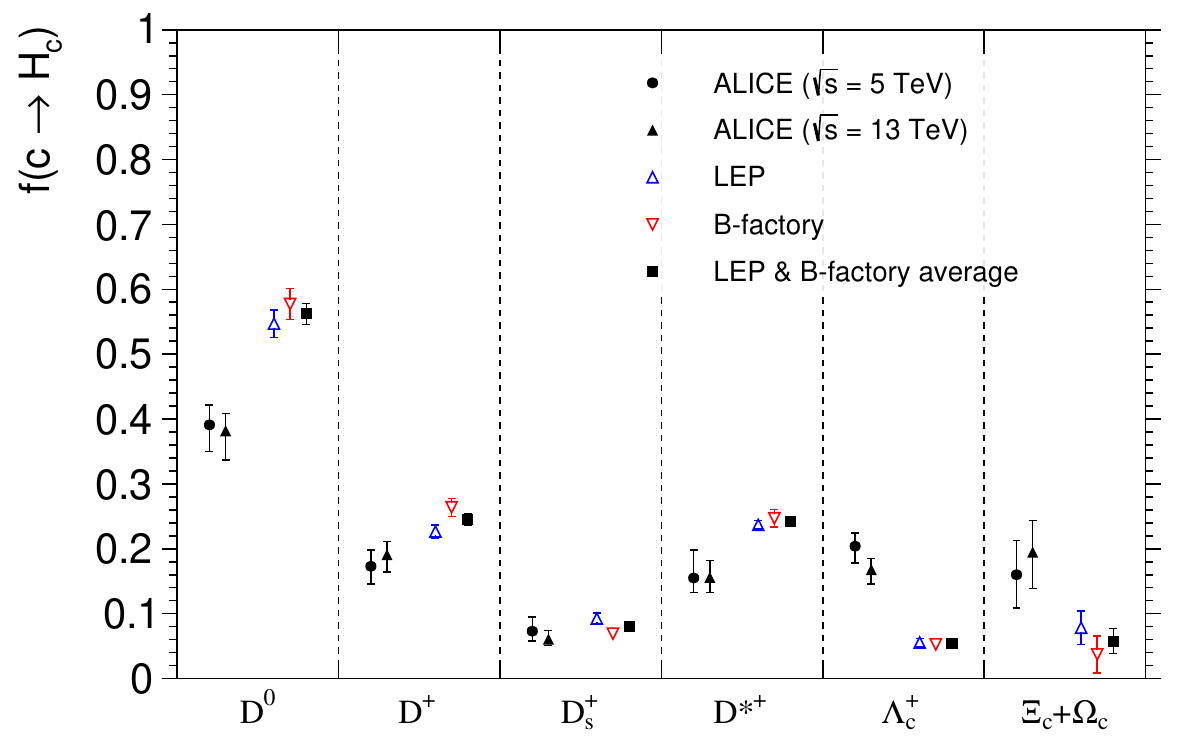}}
\end{minipage}
\hfill
\begin{minipage}{0.35\linewidth}
\centerline{\includegraphics[width=0.95\linewidth]{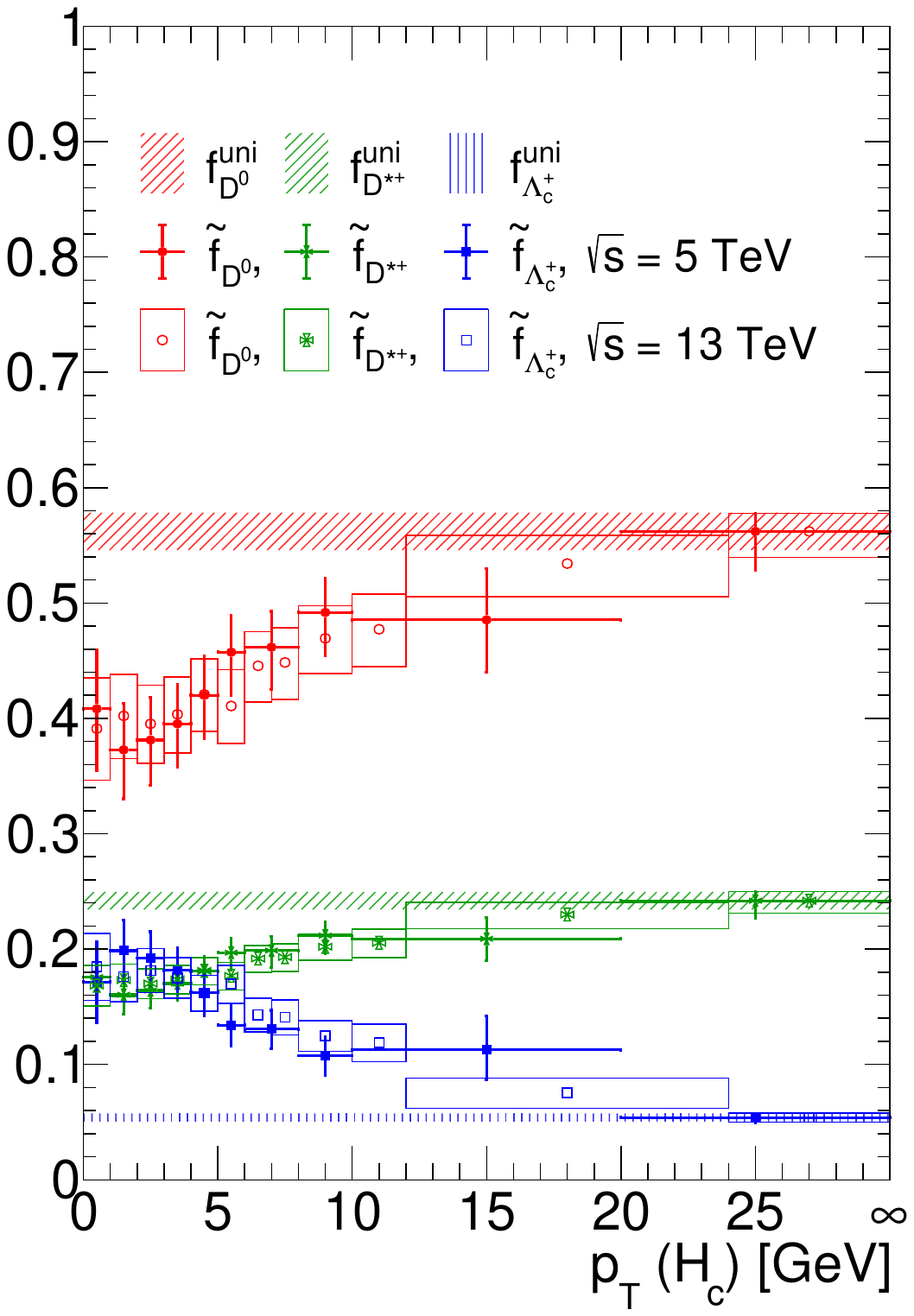}}
\end{minipage}
\caption[]{Integrated fragmentation fractions $f$ (left) and differential 
hadron fractions $\tilde f$ as a function of $p_T$ (right) at 5 and 13 TeV, 
for different charm hadron final states, compiled from\ \cite{nonuni1,nonuni2,nonuni3,nonuni4,nnloCharm1}.}
\label{fig:ftilde}
\end{figure}

Details of some theory aspects of the data driven FONLL (ddFONLL) approach
used here, previously referred to as 
``modified FONLL'', are given in a previous report\,\cite{EPS23}. In a nutshell, the 
FONLL theory, including its charm fragmentation extension, is modified
to ddFONLL by replacing the fixed ``universal'' fragmentation fraction 
$f_{H_c}^{\mathrm{uni}}$ obtained from $e^+e^-$ and/or $ep$ collisions by 
a $p_T$-dependent hadron-production 
fraction $\tilde{f}_{H_c}(p_T)$ directly obtained from measurements at LHC
(Fig. \ref{fig:ftilde}),  
\begin{equation} \label{eq:modFonll}
 d\sigma_{H_c}^{\mathrm{FONLL}} = f_{H_c}^{\mathrm{uni}} \cdot \left(d\sigma_{pp \rightarrow c}^{\mathrm{FONLL}} \otimes D_{c \rightarrow H_c}^{\mathrm{NP}}\right) \quad \to \quad d\sigma_{H_c}^{\mathrm{ddFONLL}} = \tilde{f}_{H_c}(p_T) \cdot \left(d\sigma_{pp \rightarrow c}^{\mathrm{FONLL}} \otimes D_{c \rightarrow H_c}^{\mathrm{NP}}\right)
\end{equation}
while the FONLL parametrizations of the quark-level differential 
cross sections $d\sigma_{pp \rightarrow c\bar{c}}^{\mathrm{FONLL}}$ and the 
parametrization of the nonperturbative fragmentation function 
$D_{c \rightarrow H_c}^{\mathrm{NP}}$,
here using the Kart\-velish\-vili\,\cite{Kartvelishvili} parametrization, remain 
unchanged.   
The other change is that, instead of treating the FONLL QCD parameters 
$\mu_f$ and 
$\mu_r$ (factorization and renormalization scales) and $m_c$ (charm pole mass)
as external QCD parameters, they are empirically left to float freely, 
to be fitted from data, separately for each $pp$ center-of-mass energy.  
The same is true for the Kartvelishvili parameter $\alpha_K$.
In contrast to FONLL, ddFONLL is therefore no longer a QCD theory prediction, 
rather a theory-inspired parametrization of the $pp$ data at a given 
center of mass energy. 
For the PDFs, the CTEQ6.6 set\,\cite{CTEQ66}, not relying on pp charm 
fragmentation universality as input, is used, including uncertainties 
consistent with the PROSA\cite{PROSA} low-$x$ gluon parametrization 
of the rapidity dependence (only) of LHC charm data in different regions 
of $p_T$ 
(also see previous explanations\,\cite{EPS23}).
In the limit $\tilde f(p_T) \equiv f^{\mathrm{uni}}$ (e.g. for $e^+e^-$ or $ep$), the 
standard FONLL parametrization is fully recovered, i.e. the approach remains 
fully consistent with all successful previous $e^+e^-$ and $ep$ FONLL 
applications by definition.     

\begin{figure}
\begin{minipage}{0.50\linewidth}
\centerline{\includegraphics[width=1.0\linewidth]{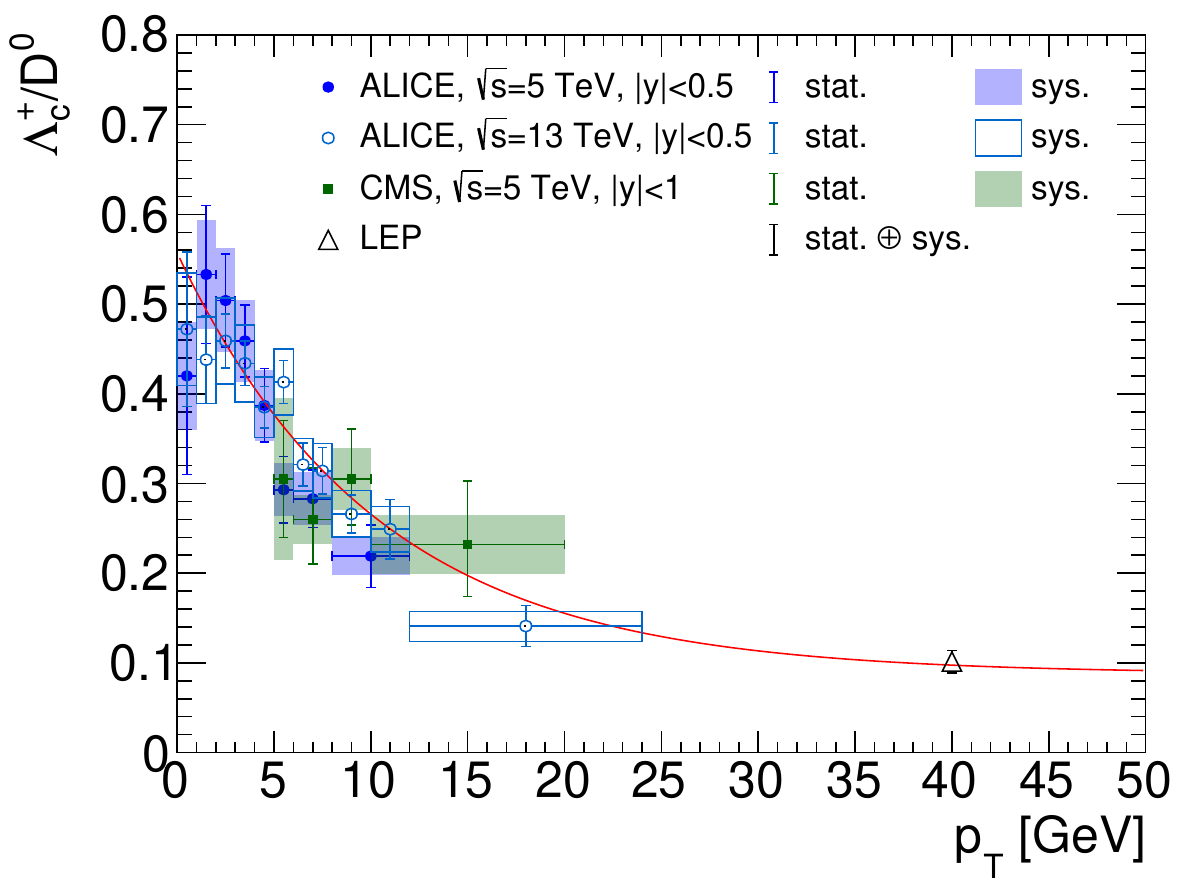}}
\end{minipage}
\hfill
\begin{minipage}{0.50\linewidth}
\centerline{\includegraphics[width=1.0\linewidth]{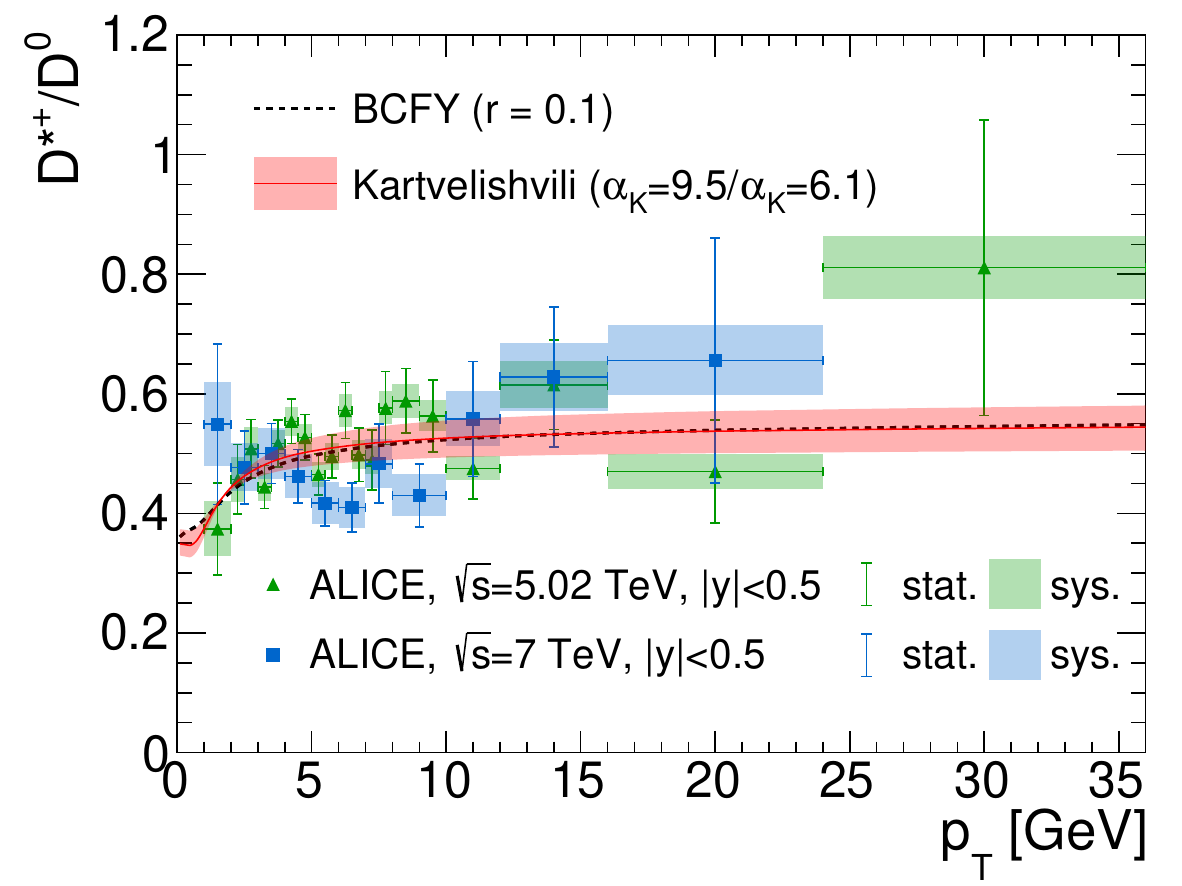}}
\end{minipage}
\hfill

\caption[]{Illustration of $p_T$ dependence and LEP asymptotic limit for the ratio 
$\Lambda_c/D^0$ (left) and $D^*/D^0$ (right). The information is compiled from \cite{nonuni2,nonuni3,nonuni4,nnloCharm1} and \cite{uni5}.}
\label{fig:LEPasymp}
\end{figure}

\begin{figure}[htbp]
\begin{minipage}{0.50\linewidth}
\centerline{\includegraphics[width=0.95\linewidth]{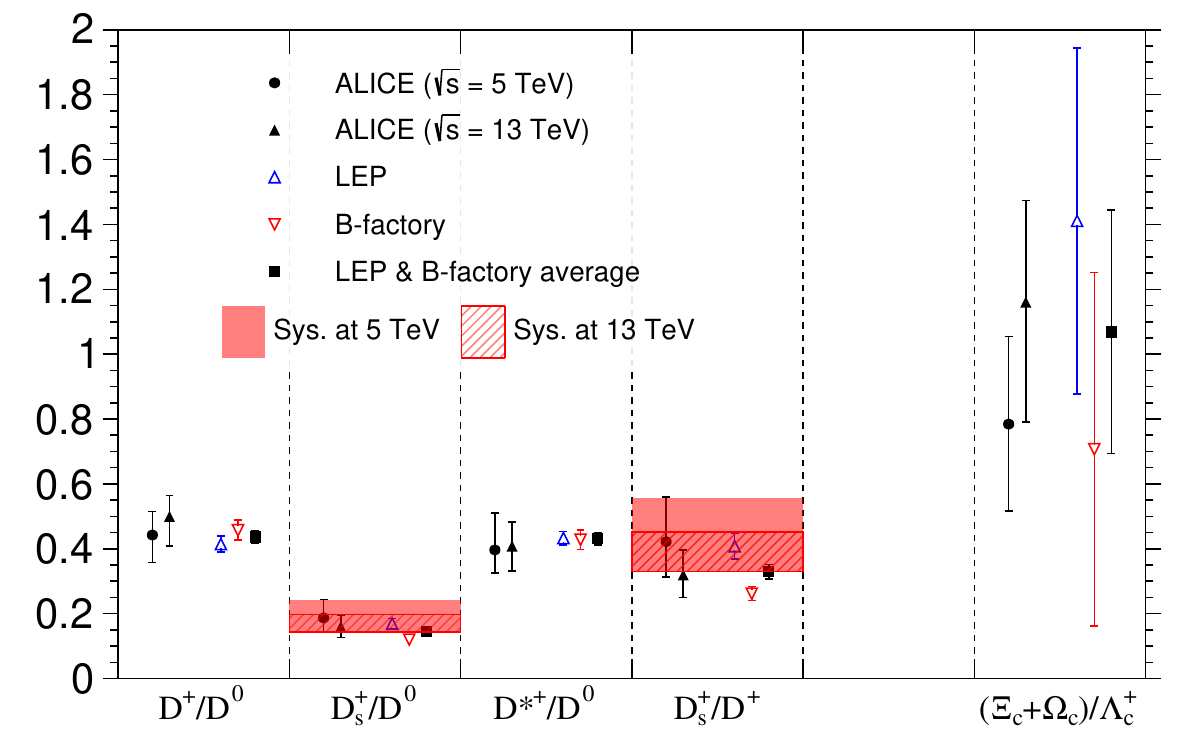}}
\end{minipage}
\hfill
\begin{minipage}{0.50\linewidth}
\centerline{\includegraphics[width=0.95\linewidth]{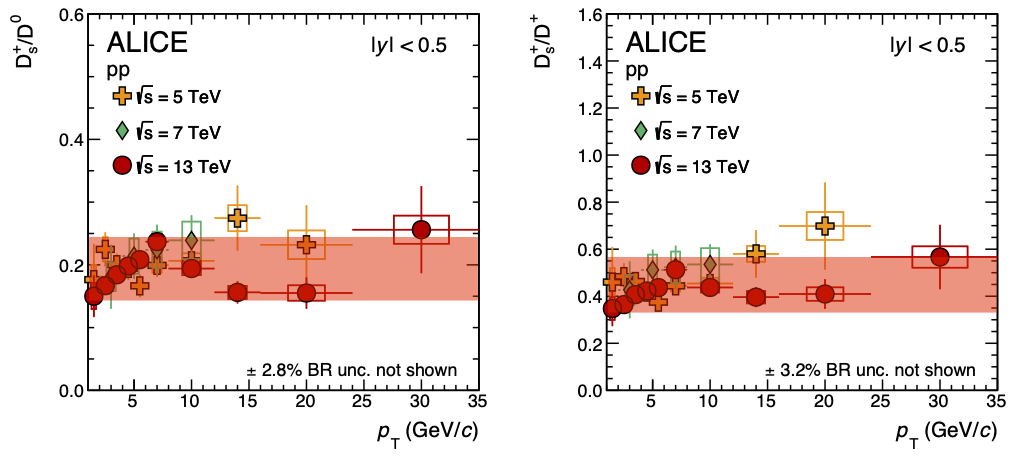}}
\end{minipage}
\caption[]{Illustration of meson-to-meson and baryon-to baryon ratios. The information is compiled from \cite{nonuni1,nonuni2,nnloCharm1}.}
\label{fig:mesmes}
\end{figure}

This ddFONLL approach implies the following assumptions: 
\begin{itemize} 
\vspace{-0.3cm}
\item The $p_T$ dependence of $\tilde f$ asymptotically approaches LEP 
$e^+e^-$ values\,\cite{HFLAV} at high $p_T$ (Fig. \ref{fig:LEPasymp}), 
i.e. at high $p_T$, charm fragmentation universality is fully recovered.  
\vspace{-0.3cm}
\item As motivated by measurements\,\cite{motrap}, there is no strong 
non-universal 
rapidity dependence of $\tilde f$, and any potential small residual dependencies
are absorbed into the global $\tilde f$ uncertainties. This assumption is also 
verified a posteriori by the fits to data below (Fig. \ref{fig:fit13TeV}) 
and elsewhere\,\cite{EPS23,DIS24}.
\vspace{-0.3cm}
\item Non-universal effects in the nonperturbative fragmentation function
$D^{NP}$ at a given $pp$ center-of-mass energy
are small enough such that the approximation of treating them as factorizable 
w.r.t. $\tilde f$ in Eq. (\ref{eq:modFonll}) ($p_T$-dependent reshuffling of 
final state charm hadron fractions) holds well enough
within current $\tilde f$ and freely fitted $\alpha_K$ uncertainties. 
This assumption is also verified a posteriori by the fits on data below.
\vspace{-0.3cm}
\item For $pp$, all ddFONLL parameters except the PDFs may be $\sqrt{s}$ dependent.   
\vspace{-0.3cm}
\end{itemize}
A further simplification made, again based on measurements, is 
\begin{itemize}
\vspace{-0.3cm}
\item While the baryon-to-meson $H_c$ ratios are strongly $p_T$ 
dependent\,\cite{nonuni2,nonuni3,nonuni4,EPS23} (Fig. \ref{fig:LEPasymp}(left)), 
the meson-to-meson and baryon-to-baryon ratios (Fig. \ref{fig:mesmes}) 
remain $p_T$ independent.
Known small deviations from this assumption, e.g. for hadrons containing 
strange vs. $u$ or $d$ quarks\,\cite{sud}, are absorbed into the global  
$\tilde f$ uncertainties. Again, this approach is also verified a posteriori
on data.   
\vspace{-0.3cm}
\end{itemize}

Finally, although not yet applied here, a check is made whether the 
parametrization above might also work for final states other than the 
weakly decaying charm ground states. 
As an example, Fig. \ref{fig:LEPasymp}(right) shows the $D^{*+}/D^0$ ratio obtained from ALICE 
for pp collisions at 5 TeV and 7 TeV, compared to a parametrization 
using different Kartvelishvili parameters for $D^*$ and $D^0$ as obtained 
from LEP data.
As expected, since part of the $D^0$ mesons originate from $D^*$ decays,
the effective $D^*$ fragmentation is a bit harder than the 
effective $D^0$ fragmentation, and the ratio can be well described by the 
LEP expectation within uncertainties, consistent with the asymptotic 
LEP universality and $D$ meson ratio universality assumptions.    
Since the Kartvelishvili parameter is a free parameter of the fit, it 
can freely adjust itself also to resonant states, if these are fitted 
``alone''. In the case of combined fits of ground states and resonant states
(not applicable here), 
some correlations between the parameters would have to be treated.  

\section{Fit Results}

\begin{figure}[htbp]
\begin{minipage}{1.0\linewidth}
 \begin{center}
 \includegraphics[width=0.16\textwidth]{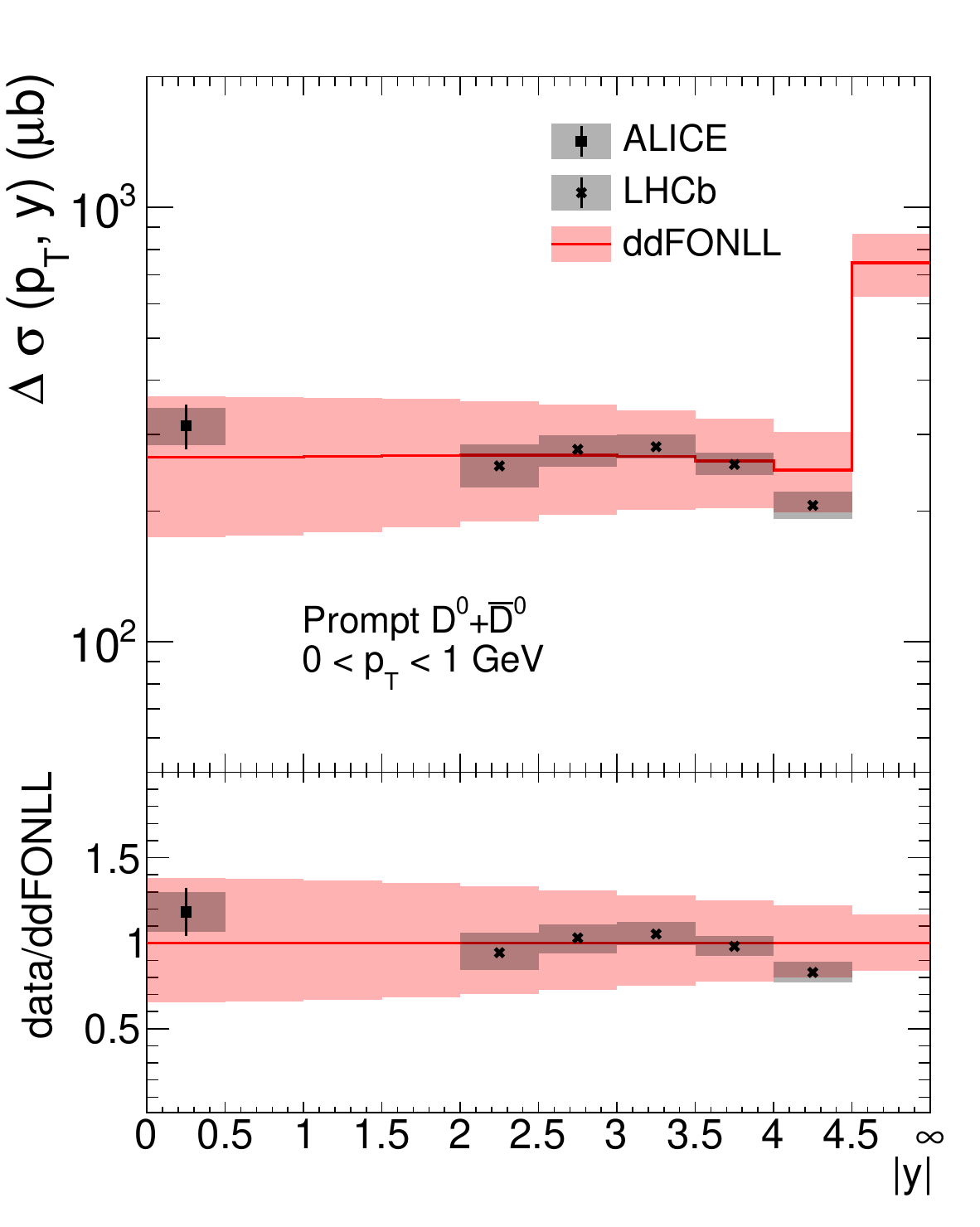}
 \includegraphics[width=0.16\textwidth]{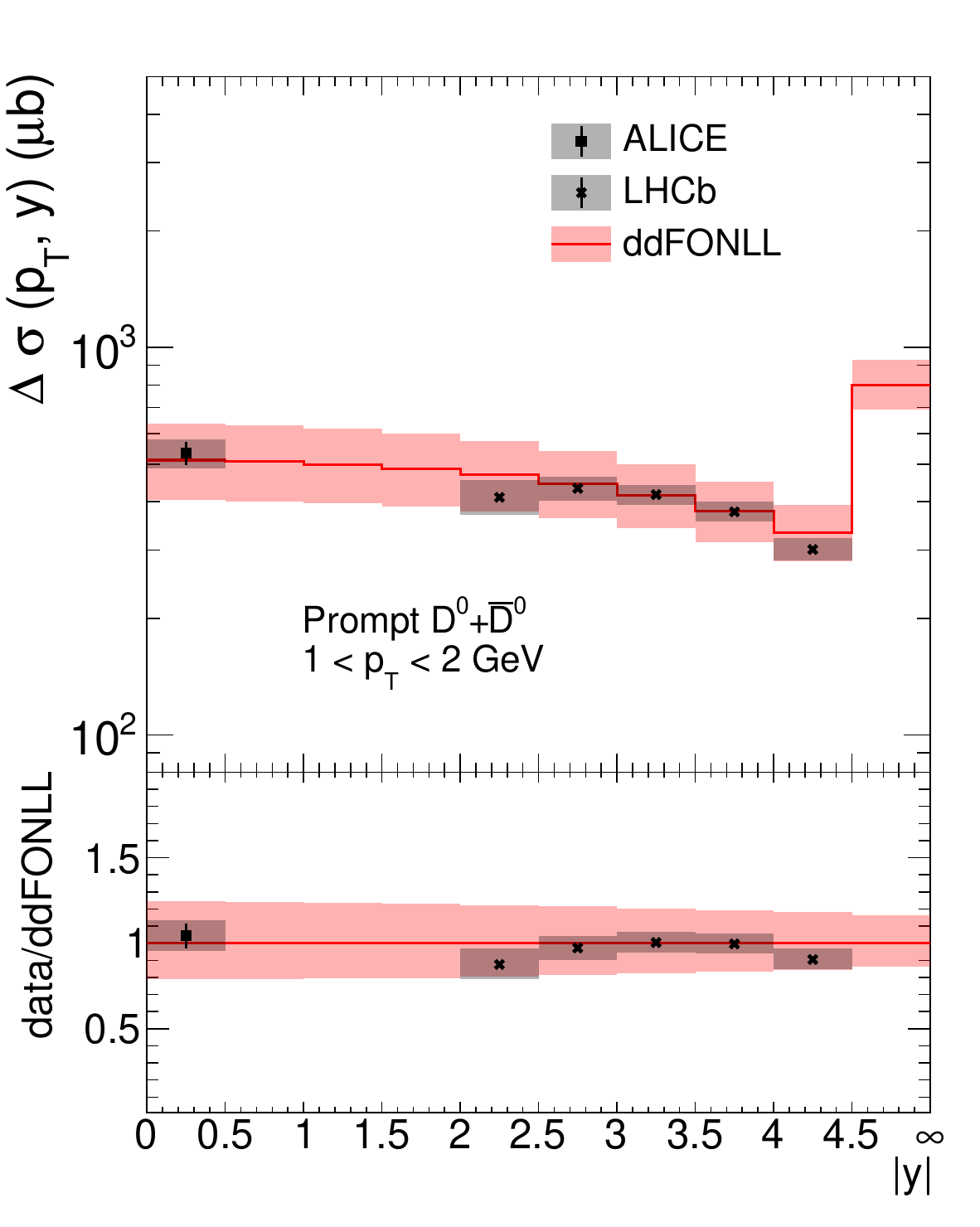}
 \includegraphics[width=0.16\textwidth]{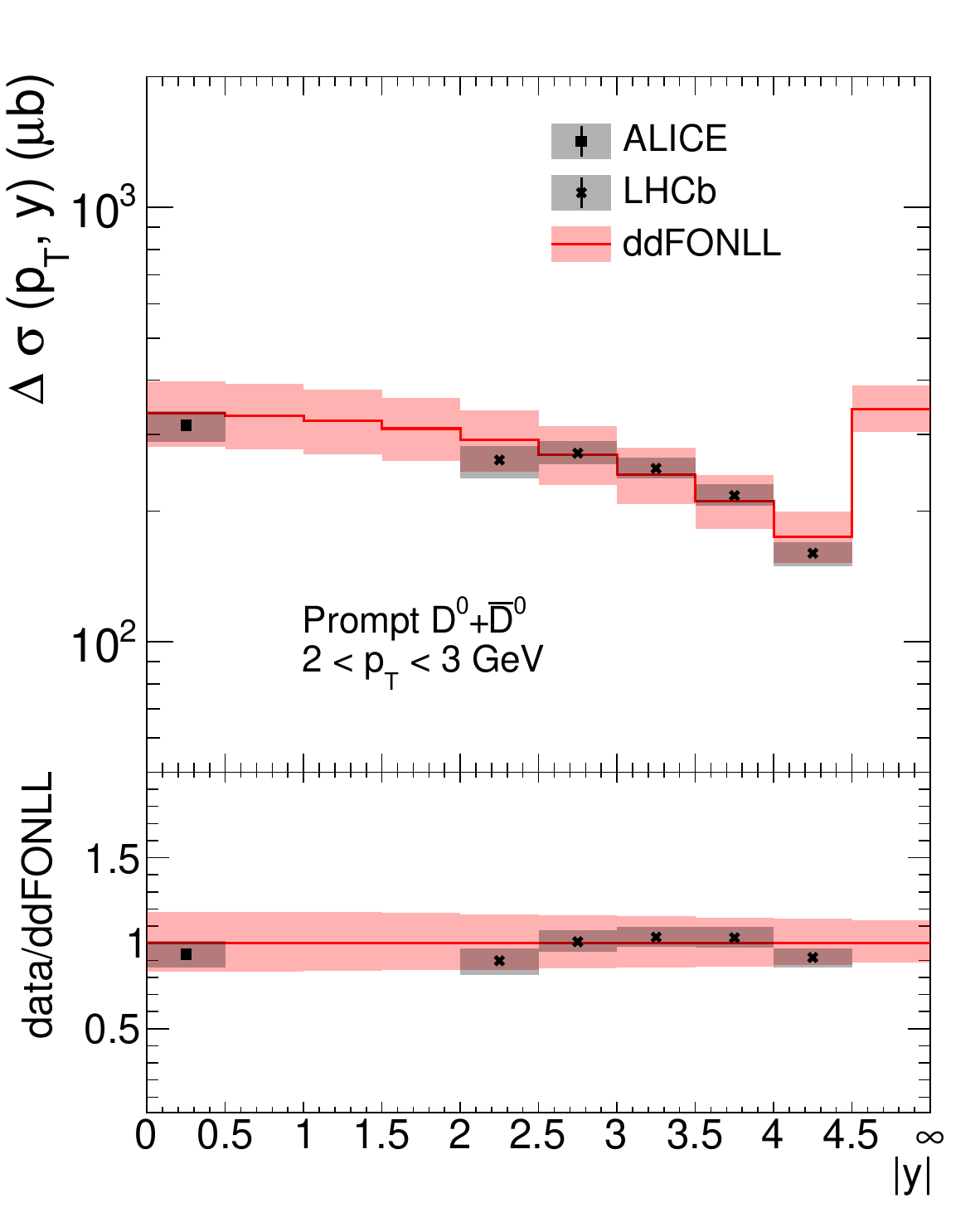}
 \includegraphics[width=0.16\textwidth]{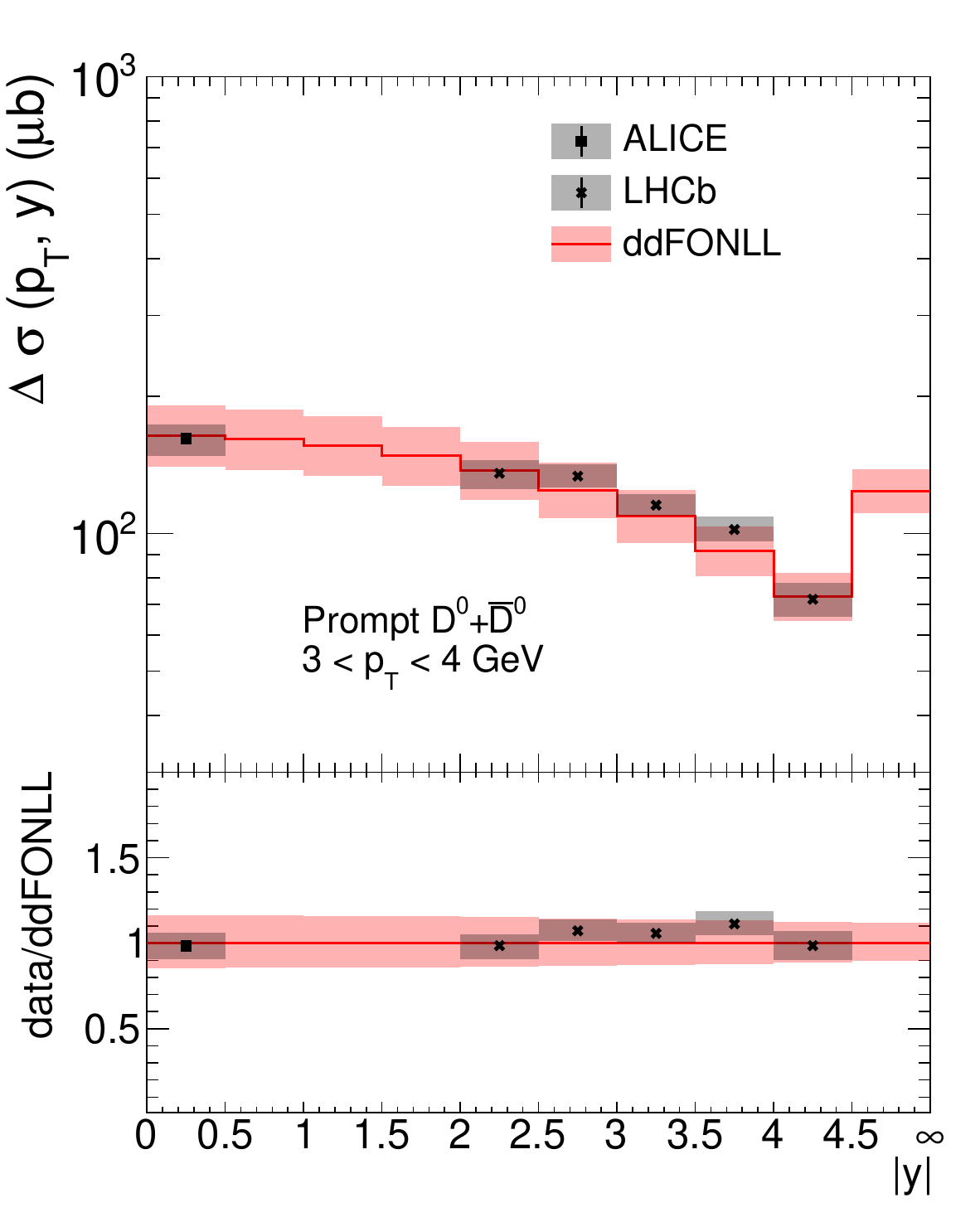}
 \includegraphics[width=0.16\textwidth]{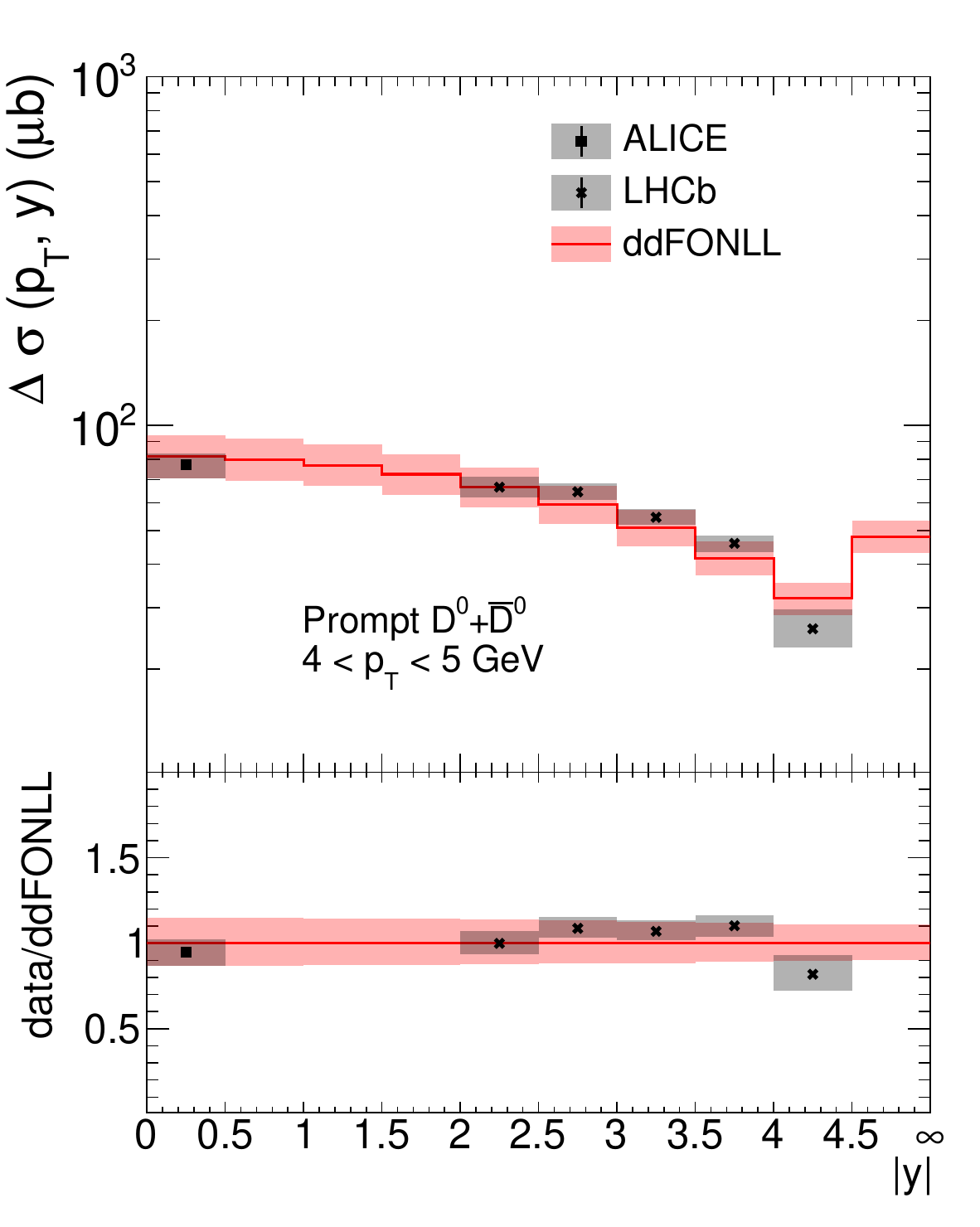}
 \includegraphics[width=0.16\textwidth]{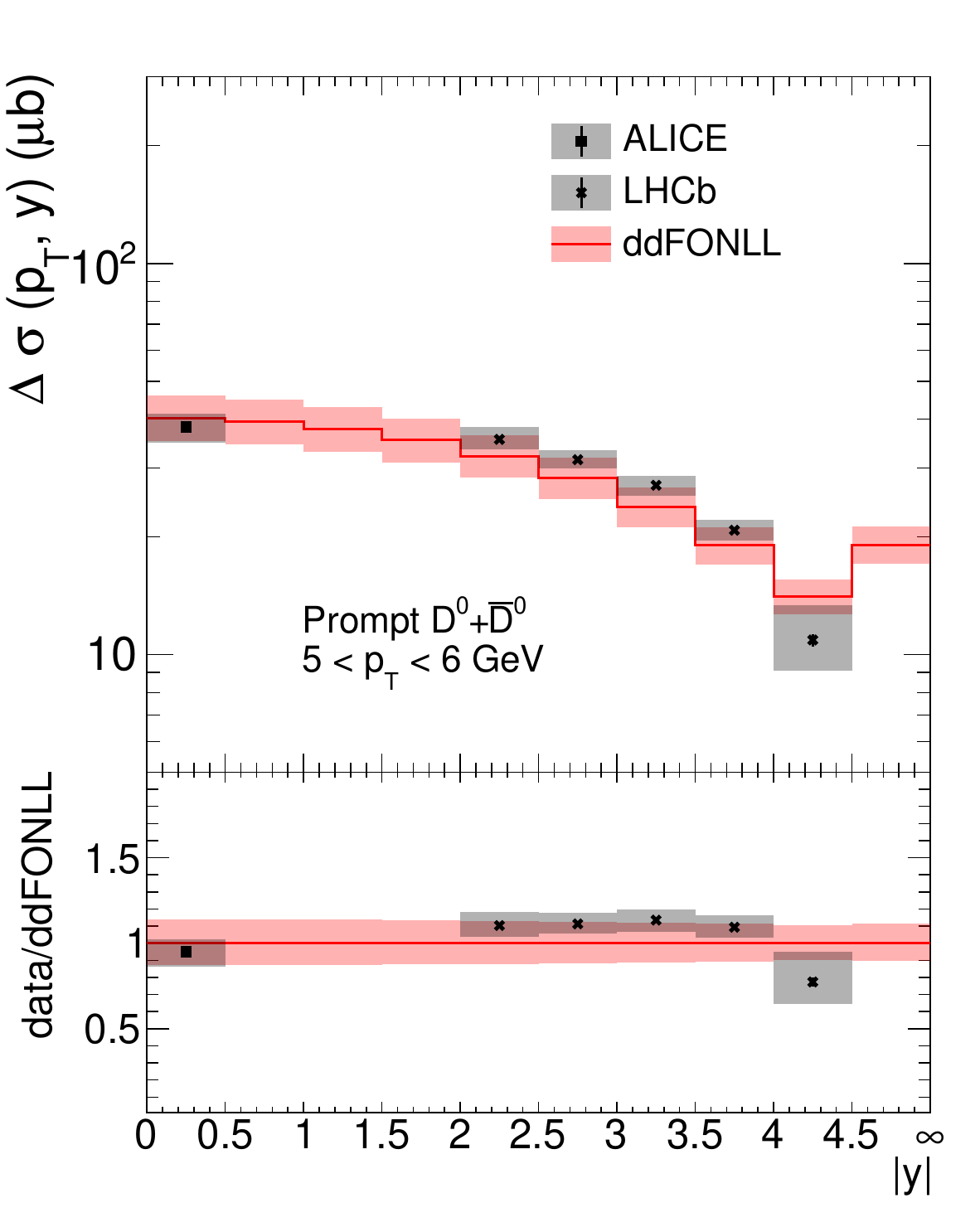}
 \includegraphics[width=0.16\textwidth]{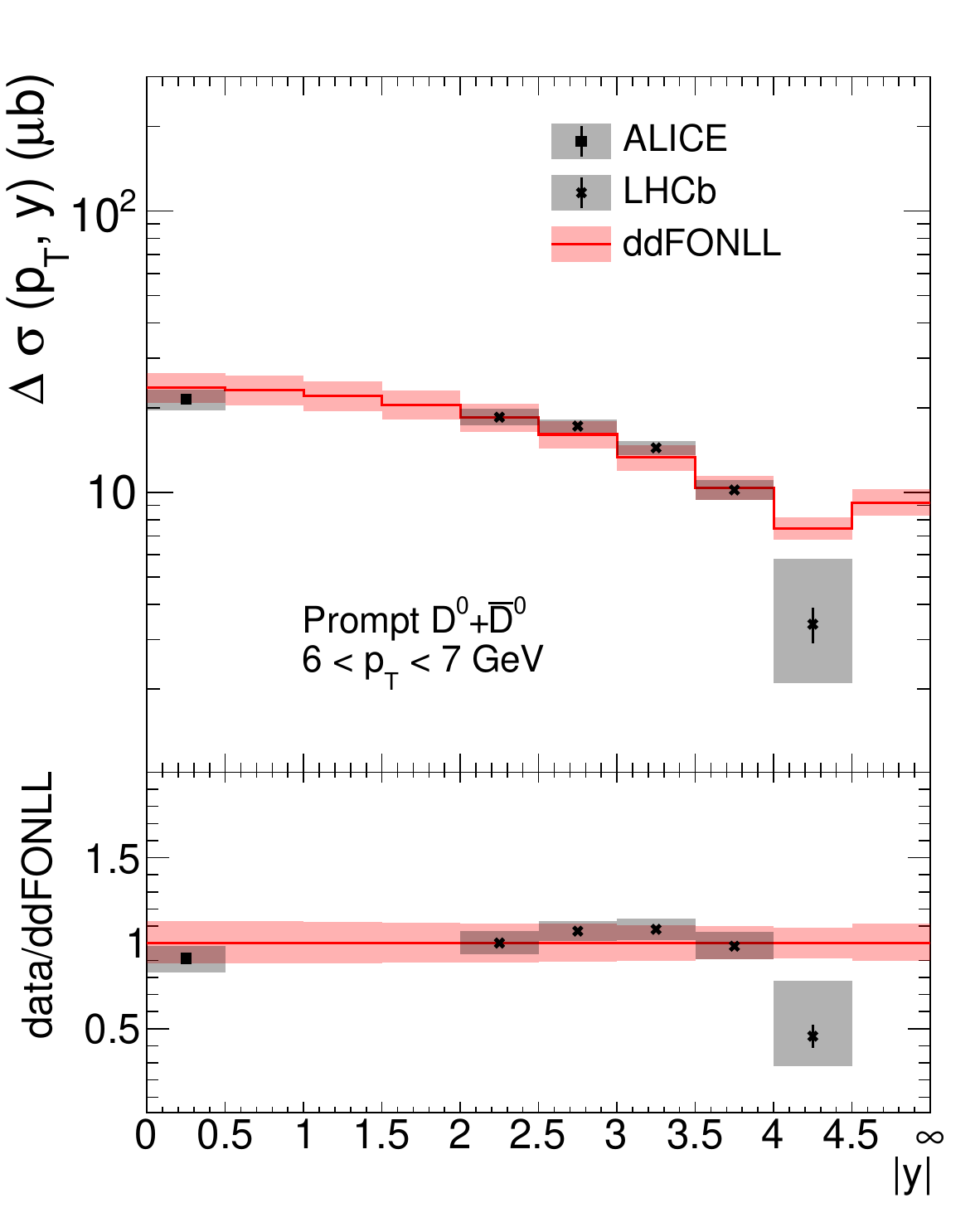}
 \includegraphics[width=0.16\textwidth]{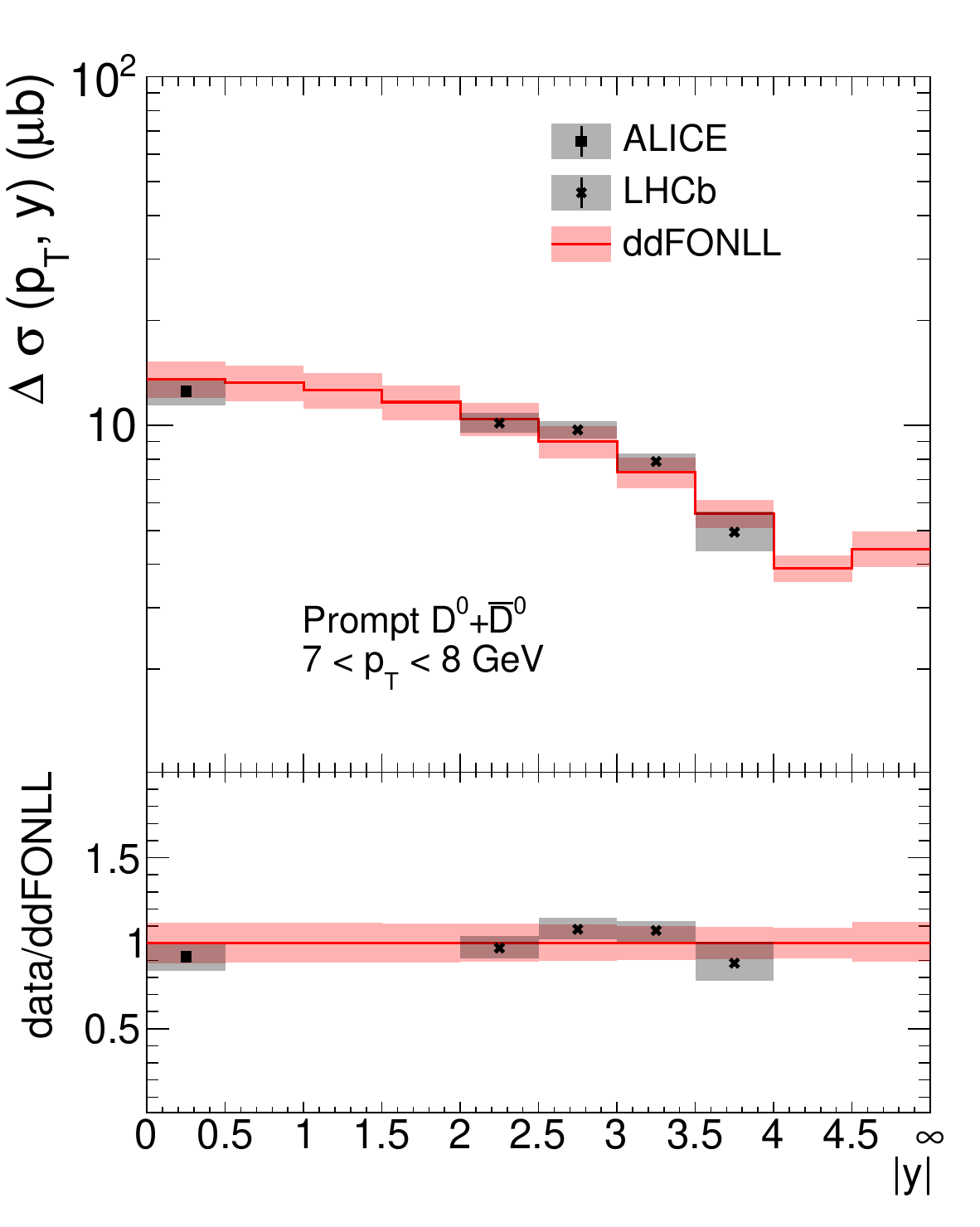}
 \includegraphics[width=0.16\textwidth]{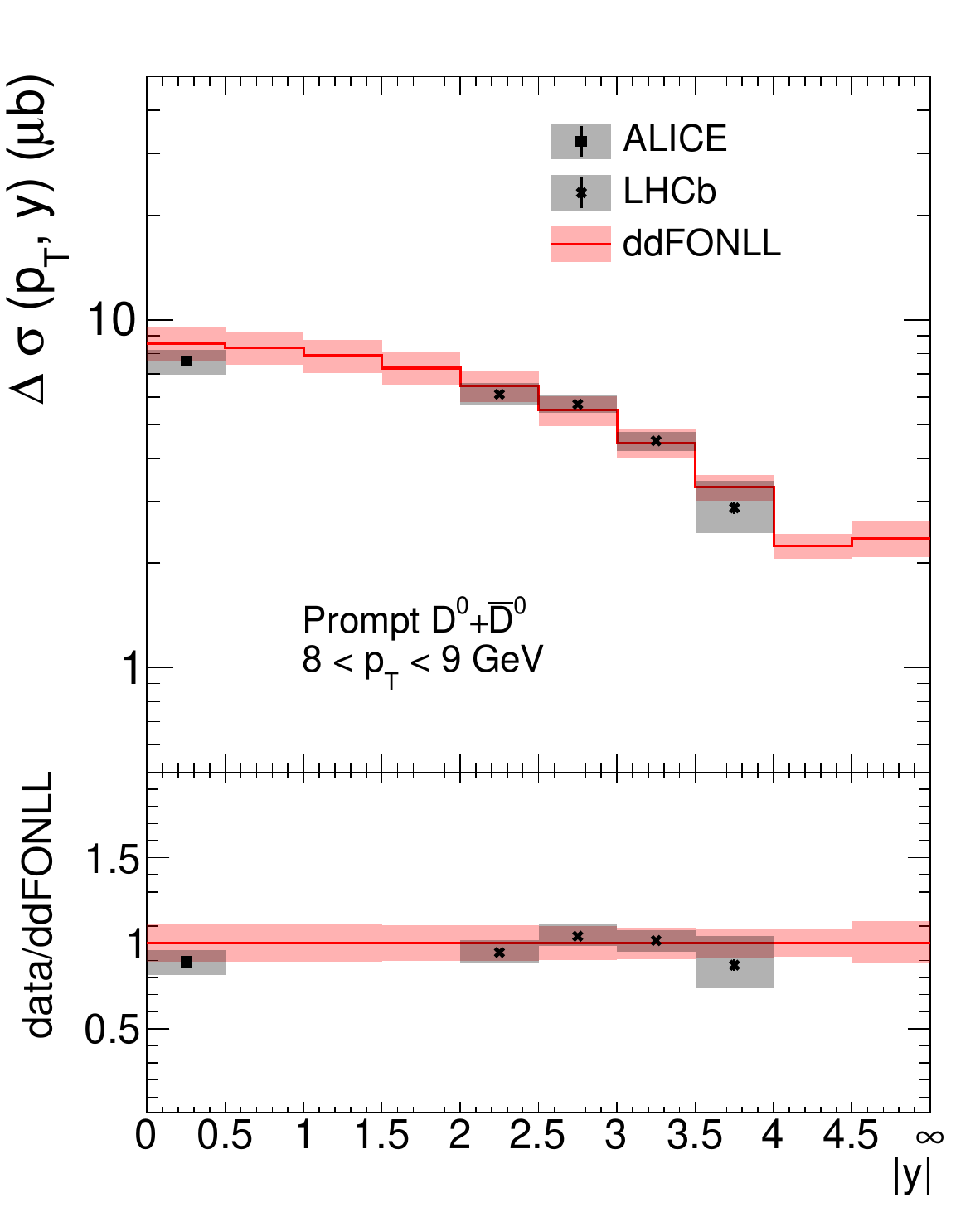}
 \includegraphics[width=0.16\textwidth]{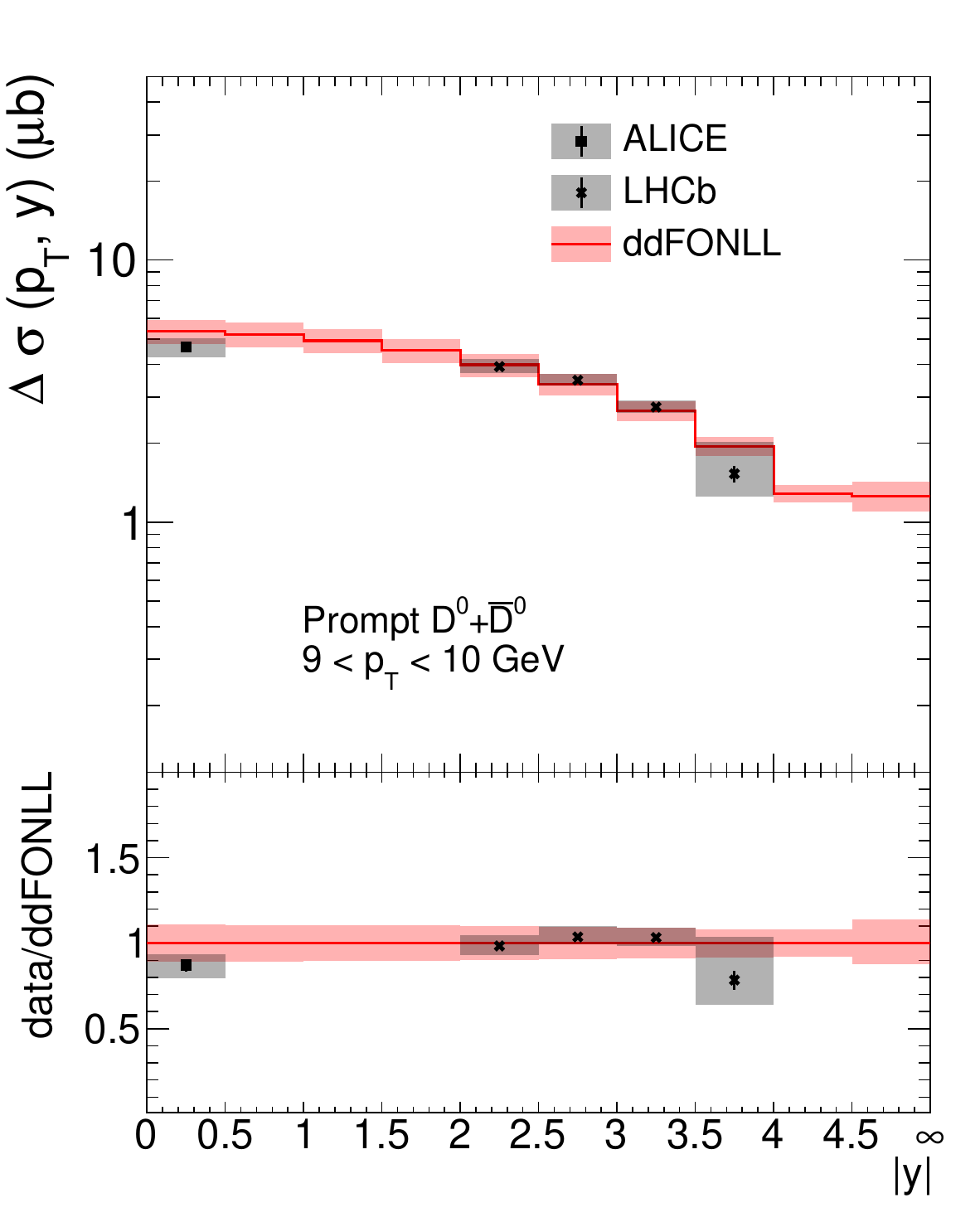}
 \includegraphics[width=0.16\textwidth]{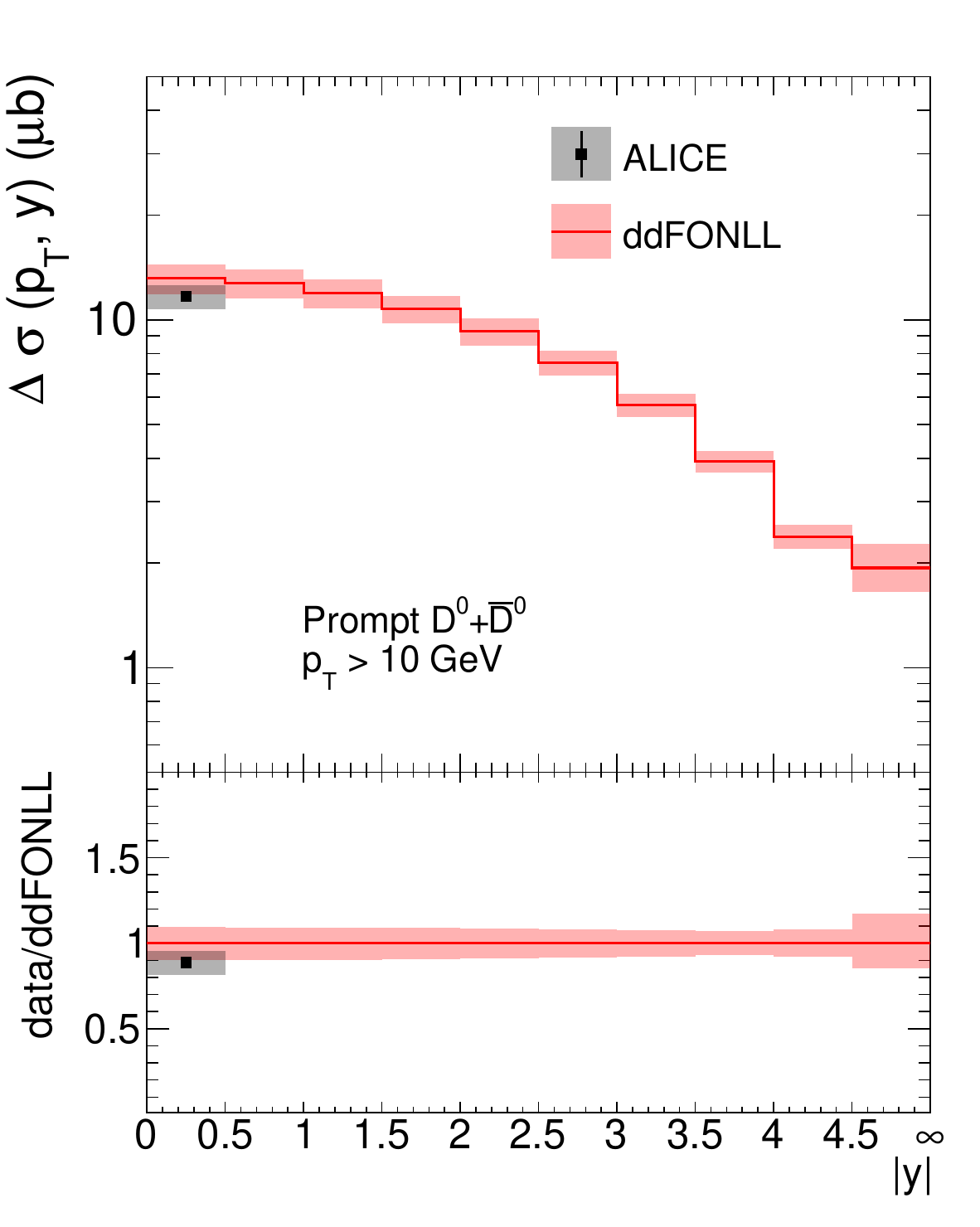}
 \end{center}
\end{minipage}
\caption[]{$D^0+\overline{D}{}^0$ cross sections as a function $|y|$ in bins 
of $p_T$ for $\sqrt{s}=13$ TeV. The red bands are the data driven FONLL as 
obtained with non-universal charm fragmentation, which describe the 
data\,\cite{nonuni2,uni6} 
(black points/grey boxes) well in the full phase space. 
The total uncertainty of the data driven FONLL includes the uncertainties 
of the CTEQ6.6 PDF\,\cite{CTEQ66}, 
$\tilde f$ (Fig. \ref{fig:ftilde}),
and the $\chi^2$ scan.} 
\label{fig:fit13TeV}
\end{figure}

\begin{figure}[htbp]
\begin{minipage}{0.45\linewidth}
\centerline{\includegraphics[width=0.95\linewidth]{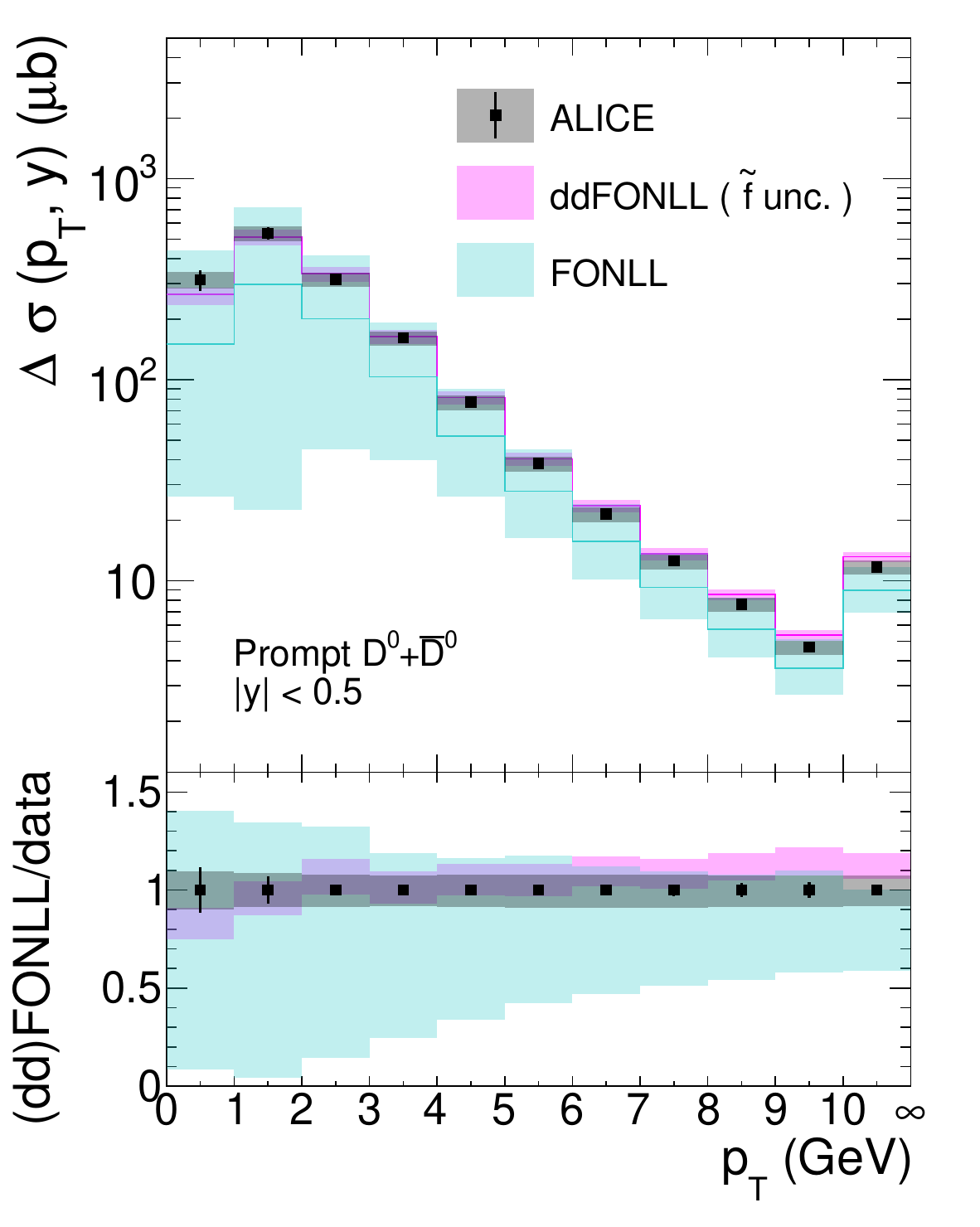}}
\end{minipage}
\begin{minipage}{0.45\linewidth}
\centerline{\includegraphics[width=0.95\linewidth]{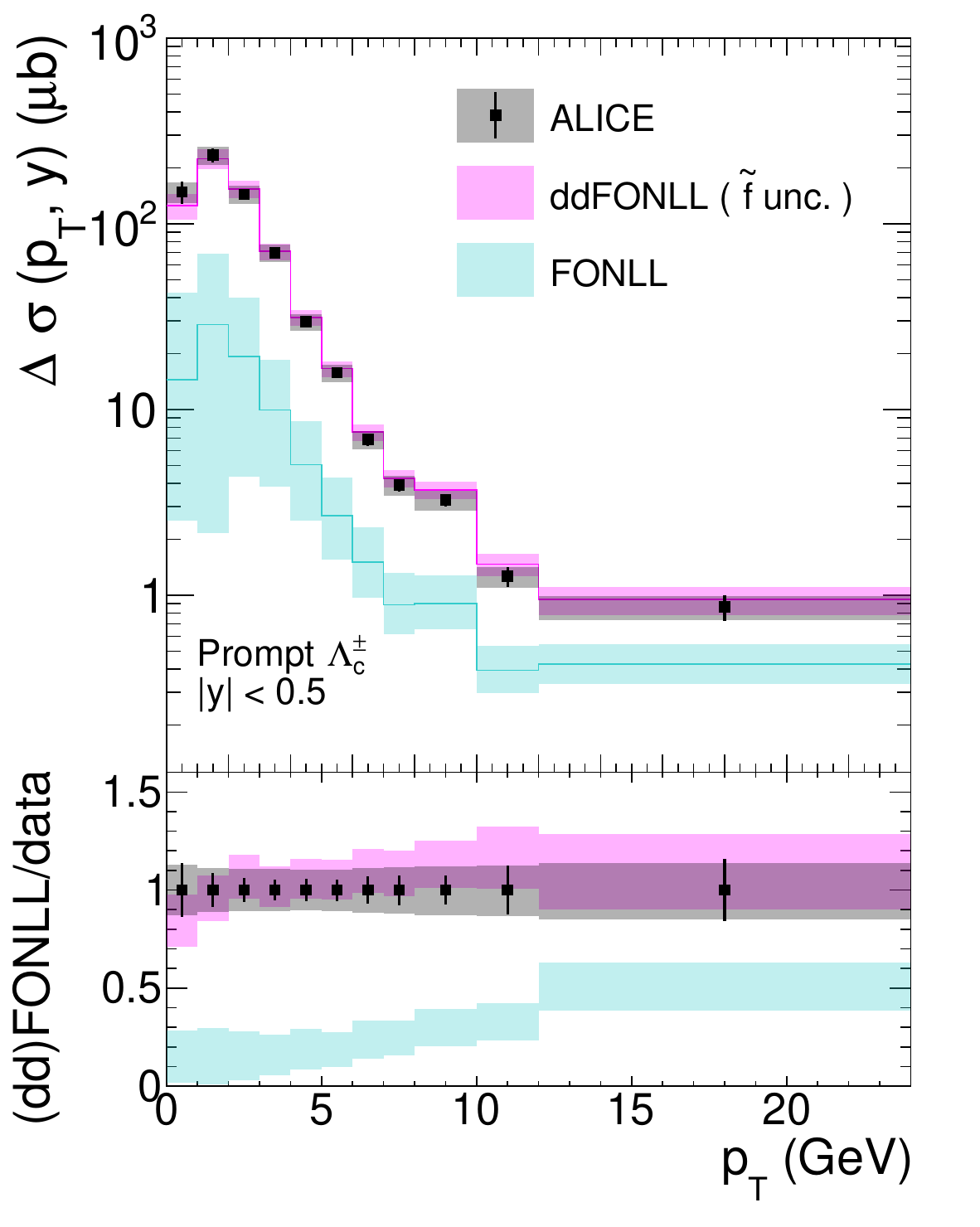}}
\end{minipage}
\caption[]{Result of the fitted 13 TeV ddFONLL parametrization for $D^0$ (left) and $\Lambda_c$ (right), compared to ALICE data\,\cite{nonuni2} and to the standard FONLL prediction. Only the $\tilde f$ uncertainties are shown 
here for the ddFONLL case.}
\label{fig:Lambdac13TeV}
\end{figure}

The results of fitting the free ddFONLL parameters $\mu_f$, $\mu_r$, $m_c$
and $\alpha_K$ to 5 TeV $D^0$ data were already shown previously\,\cite{EPS23}. 
These results remain unchanged. The results of fitting these parameters
to 13 TeV $D^0$ data are shown in Fig. \ref{fig:fit13TeV}. 
Again, very good agreement with 
data is obtained, reflected by the $\chi^2$ S-factor for the 4D fit coming 
out very close to 1. The fit parameters, which will be discussed further in 
another document\,\cite{DIS24}, 
are consistent with those obtained at 5 TeV within uncertainties.
In both cases, the fitted $\alpha_K$ parameters are also consistent with those
expected from the asymptotic agreement with LEP (see above). 
For illustration, the result of the 13 TeV ddFONLL parametrization obtained 
from the fit is also shown in Fig. \ref{fig:Lambdac13TeV} 
for the comparison to ALICE $D^0$ and $\Lambda_c$
assuming charm fragmentation universality. Both the shape and normalization
of both distributions are well described by ddFONLL while standard FONLL fails 
in both shape and normalization for the $\Lambda_c$ case. 
A very similar comparison is also available for the 
5 TeV case\,\cite{EPS23,DIS24}. 
These results are fully consistent with all the explicit and implicit 
ddFONLL assumptions stated above (a posteriori verification). 

\begin{figure}[htbp]
\begin{minipage}{0.70\linewidth}
\centerline{\includegraphics[width=0.95\linewidth]{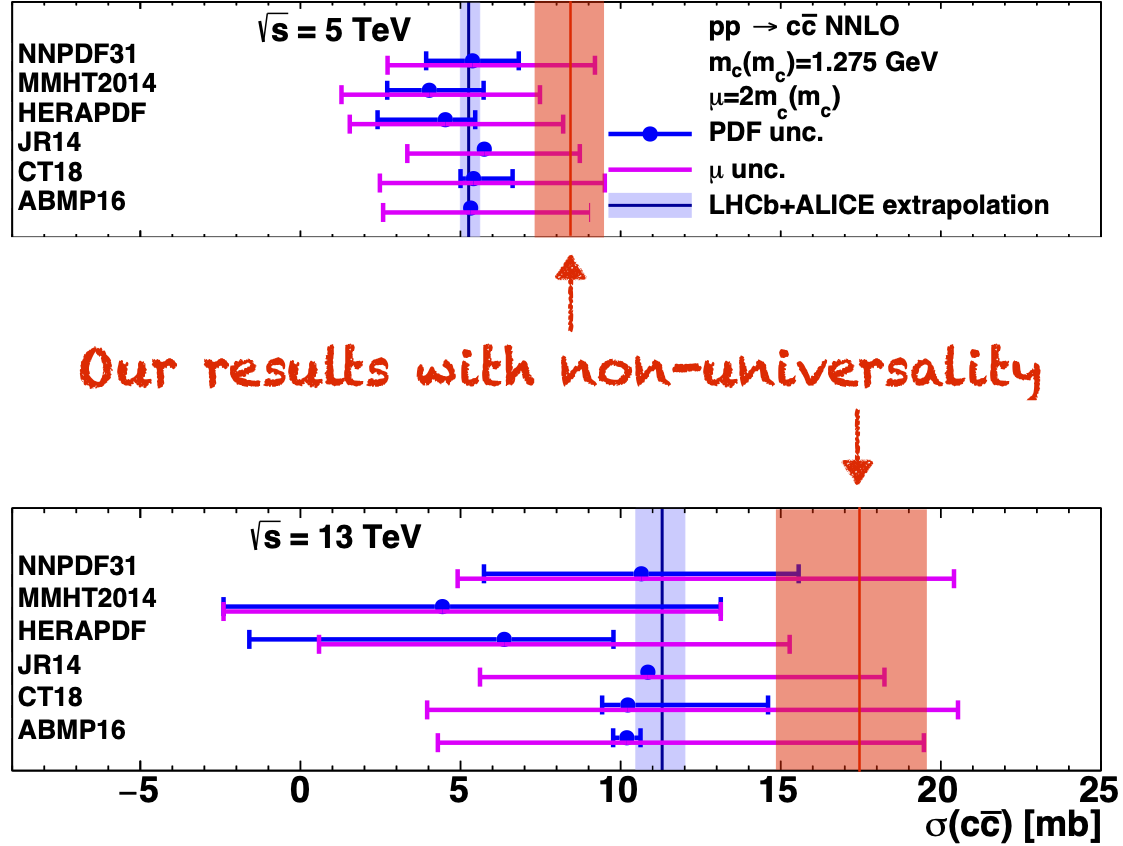}}
\end{minipage}
\caption[]{Comparison of measured total charm-pair cross sections at 
5 TeV and 13 TeV (red bands) to previous determinations\,\cite{nnloCharm1} 
(blue bands) and to NNLO predictions with various PDF sets including 
theory uncertainties (points with bars). Plots adapted from \cite{nnloCharm1}.}
\label{fig:NNLOcomp}
\end{figure}

Using the measurements for all bins in which measurements are available,
and the ddFONLL data parametrization in all others,  
the resulting 5 TeV and 13 TeV total charm pair production cross sections 
are obtained to be 
\begin{eqnarray}
 \sigma^{\mathrm{tot}}_{c\bar{c}}(5\ \mathrm{TeV}) &
= \ \ 8.43 ^{+0.25}_{-0.25}\mathrm{(data)} ^{+0.40}_{-0.42}(\tilde{f}) ^{+0.67}_{-0.56}\mathrm{(PDF)} ^{+0.13}_{-0.12}(\mu_f, \mu_r, m_c, \alpha_K) ^{+0.65}_{-0.88}(f^{pp}_{D^0}) \mathrm{[mb]} \\
 & = \ \ 8.43 ^{+1.05}_{-1.16}(\mathrm{total})\ \mathrm{mb}. \hspace{8.2cm} \\
 \sigma^{\mathrm{tot}}_{c\bar{c}}(13\ \mathrm{TeV}) &
= 17.43 ^{+0.56}_{-0.53}\mathrm{(data)} ^{+0.69}_{-0.78}(\tilde{f}) ^{+1.47}_{-1.22}\mathrm{(PDF)} ^{+0.24}_{-0.18}(\mu_f, \mu_r, m_c, \alpha_K) ^{+1.19}_{-2.05}(f^{pp}_{D^0}) \mathrm{[mb]} \\
 & = 17.43 ^{+2.10}_{-2.57}(\mathrm{total})\ \mathrm{mb}. \hspace{8.2cm}
\end{eqnarray}
in which $f^{pp}_{D^0}$ refers to the integrated $D^0$ fragmentation fraction
measured at 5 TeV or 13 TeV, respectively. The respective extrapolation 
factors for unmeasured phase space are about 1.8 and 1.9. 
In Fig. \ref{fig:NNLOcomp} these total cross sections are compared to previous 
determinations\,\cite{nnloCharm1}  
still based on the charm fragmentation universality assumption, and to 
NNLO predictions. The complete treatment of charm fragmentation nonuniversality 
significantly increases the extracted total charm cross sections, and therefore 
replaces all previous such determinations. 
The measurements are still consistent with the NNLO predictions, but now 
situated towards the upper edge of the NNLO theory uncertainty band.     

Very preliminary studies of the sensitivity of these predictions to the 
MSbar charm quark mass, further detailed elsewhere\,\cite{DIS24}, 
indicate that the mass preferred by the data is 
consistent with the world average and can be constrained with an 
uncertainty of order 200 MeV by these measurements. This uncertainty is 
still significantly larger than corresponding constraints e.g. from QCD 
fits of $ep$ data\,\cite{charmmassep}, 
but a first step towards such a constraint from $pp$ data at NNLO, and at least
a very important consistency check. Furthermore, depending on the stiffness
of the low-$x$ PDF parametrization, these same studies indicate a significant 
potential for contraining the gluon distribution at very low $x$ at NNLO, 
in an $x$ region where data constraints\,\cite{PROSA} were so far possible only 
at NLO.   
More details on these studies will be given elsewhere\,\cite{DIS24}. 
 
%

\end{document}